\documentclass[epj]{svjour}
\usepackage{graphics}
\begin{document}
\title{A Realistic Solvable Model for the Coulomb Dissociation of
Neutron Halo Nuclei}
\author{Gerhard Baur\inst{1} \and \underline{Kai Hencken}\inst{2}
\and Dirk Trautmann\inst{2}
}                     
\institute{Institut f\"ur Kernphysik (Theorie), Forschungszentrum J\"ulich,
52425 J\"ulich, Germany
\and 
Institut f\"ur Physik, Universit\"at Basel, Klingelbergstr. 82,
4056 Basel, Switzerland}
\date{Received: \today / Revised version: \today}
%
\abstract{
As a model of a neutron halo nucleus we consider a neutron bound to an
inert core by a zero range force. We study the breakup of this simple
nucleus in the Coulomb field of a target nucleus. In the post-form
DWBA (or, in our simple model CWBA (``Coulomb Wave Born
Approximation'')) an analytic solution for the $T$-matrix is known. We
study limiting cases of this $T$-matrix. As it should be, we recover
the Born approximation for weak Coulomb fields (i.e., for the relevant
Coulomb parameters much smaller than 1). For strong Coulomb fields,
high beam energies, and scattering to the forward region we find a
result which is very similar to the Born result.  It is only modified
by a relative phase (close to 0) between the two terms and a prefactor
(close to 1). A similar situation exists for bremsstrahlung
emission. This formula can be related to the first order semiclassical
treatment of the electromagnetic dissociation. Since our CWBA model
contains the electromagnetic interaction between the core and the
target nucleus to all orders, this means that higher order effects
(including postacceleration effects) are small in the case of high
beam energies and forward scattering. Our model also predicts a
scaling behavior of the differential cross section, that is, different
systems (with different binding energies, beam energies and scattering
angles) show the same dependence on two variables $x$ and $y$.
} 
\maketitle
\section{Introduction}
\label{intro}
Breakup processes in nucleus-nucleus collisions are complicated, in
whatever way they are treated. They constitute at least a three-body
problem, which is further complicated due to the long range Coulomb
force.  Exact treatments (like the Faddeev-approach) are therefore
prohibitively cumbersome.  On the other hand, many approximate schemes
have been developed in the field of direct nuclear reactions, and
these approaches have been used with considerable success
\cite{Austern70}. In this context we wish to investigate a realistic
model for the Coulomb breakup of a neutron halo nucleus. With the
operation of exotic beam facilities all over the world, these
reactions (previously restricted essentially to deuteron induced
reactions) have come into focus again. The Coulomb breakup of these
nuclei is of interest also for nuclear astrophysics, since the breakup
cross section can be related to the photo-dissociation cross section
and to radiative capture reactions relevant for nuclear astrophysics
\cite{BaurR96}.

An important benefit of the present model is that it can be solved
analytically in the DWBA (CWBA) approximation.  Thus it constitutes an
ideal ``theoretical laboratory'' to investigate the physics of breakup
reactions, certain limiting cases and its relation to other models
like the semiclassical approximation, which is mainly used in the
interpretation of experiments. Especially the effect of
postacceleration (to be explained in more detail below) can be studied
in a unique way in this approach.

\section{Description of Theoretical Model}
\label{sec:1}

We consider the breakup of a particle $a=(c+n)$ (deuteron,
neutron-halo nucleus) consisting of a loosely bound neutral particle
$n$ and the core $c$ (with charge $Z_c$) in the Coulomb field of a
target nucleus with charge $Z$.
\begin{equation}
a+Z \rightarrow c+n+Z.
\label{eq:1}
\end{equation}
As a further simplification the $a=(c+n)$ system is assumed to be
bound by a zero range force. The bound-state wave function of the
system is given by
\begin{equation}
\phi_0 = \sqrt{\frac{\kappa}{2\pi}} \frac{\exp(-\kappa r)}{r},
\end{equation}
where the quantity $\kappa$ is related to the binding energy
$E_{\mathrm{bind}}$ of the system by
\begin{equation}
E_{\mathrm{bind}} = \frac{\hbar^2 \kappa^2}{2 \mu},
\qquad \mu = \frac{m_n m_c}{m_n+m_c}.
\end{equation}

In the post-form CWBA the T-matrix for the reaction Eq.~(\ref{eq:1})
can be written as \cite{BaurT72}
\begin{eqnarray}
T &=& \left< \chi^{(-)}_{\vec q_c} \psi_{\vec q_n} \right| V_{nc}
\left| \chi_{\vec q_a}^{(+)} \phi_0 \right>
\label{eq:postdwba}\\
&=& D_0 \int d^3R \chi^{(-)}_{\vec q_c}(\vec R) e^{-i \vec q_n\cdot\vec R}
\chi_{\vec q_a}^{(+)}(\vec R),
\label{eq:tintegral}
\end{eqnarray}
with the ``zero range constant'' $D_0$ given by $D_0 =
\frac{\hbar^2}{2 \mu} \sqrt{8 \pi \kappa}$.  The initial state is
given by the incoming Coulomb wave function $\chi^{(+)}_{\vec q_a}$
with momentum $\vec q_a$ and the halo wave function $\phi_0$. The
final state is given by the independent motion of the core described
by the outgoing Coulomb wave function $\chi^{(-)}_{\vec q_c}$ in the
Coulomb field of the target nucleus $Z$ with asymptotic momentum $\vec
q_c$ and the free neutron with momentum $\vec q_n$, described by a
plane wave. In these wave functions the Coulomb interaction is taken
into account correctly to all orders.  In our model there is no
resonance structure in the $c+n$ continuum.  This is clearly a good
assumption for the deuteron and also for other neutron halo systems.

There exists another form of the $T$-matrix element, which is not
equivalent to Eq.~(\ref{eq:tintegral}). It is called the
``prior-form'' \cite{Austern70}.  The final state is described by a
c.m. motion of the $(c+n)$ system (as a Coulomb wave function) and a
relative wave function of the unbound $(c+n)$ system.

\begin{figure}[tb]
\begin{center}
\resizebox{0.30\textwidth}{!}{%
\includegraphics{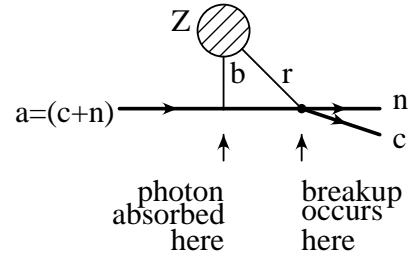}}
\end{center}
\caption{A schematic view of Coulomb-breakup, as adapted from
\protect\cite{Galonsky94}. The distance from the target nucleus to the
breakup point is denoted by $r$.}
\label{fig:1}
\end{figure}

\begin{figure}[tb]
\begin{center}
\resizebox{0.45\textwidth}{!}{%
\includegraphics{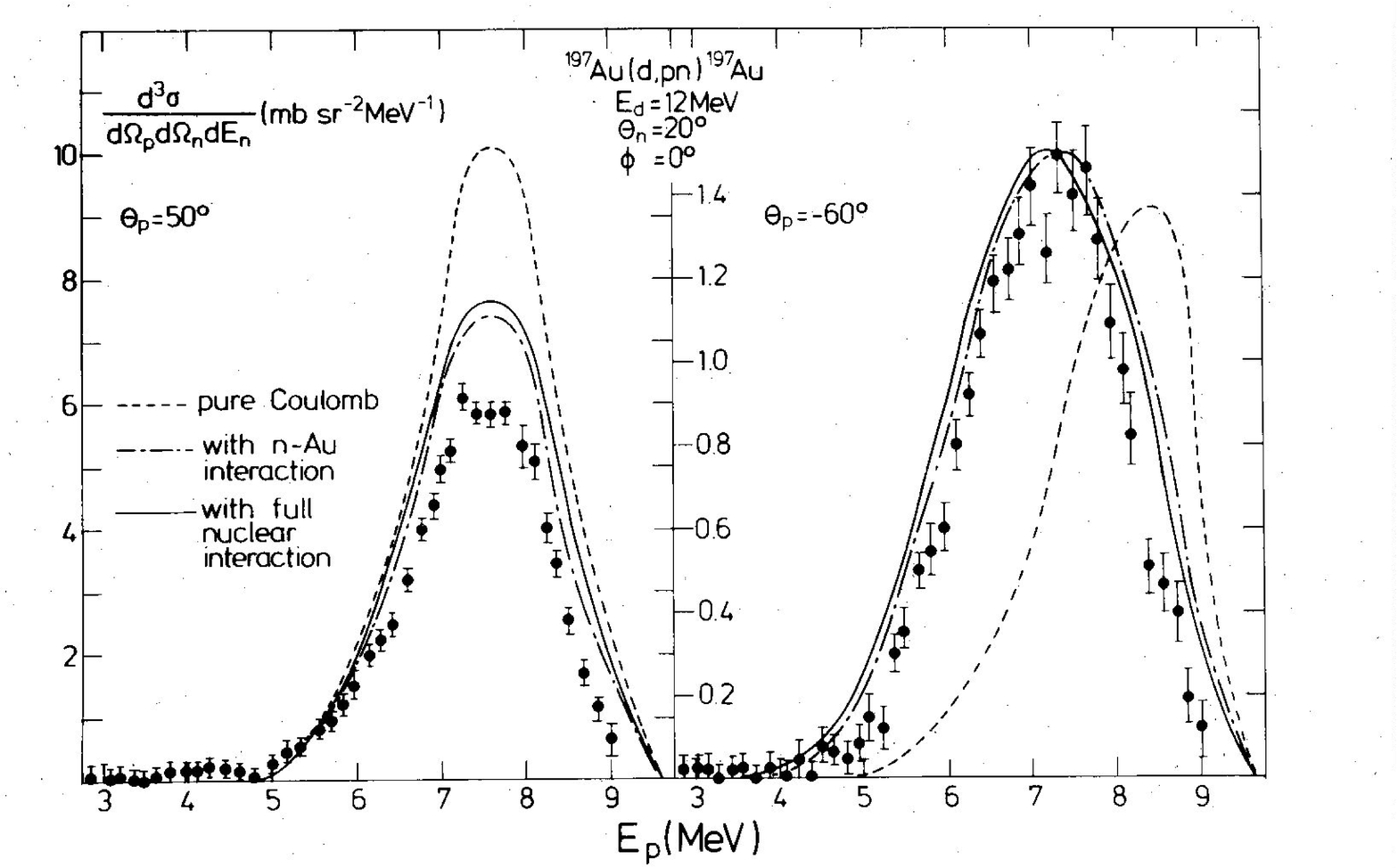}}
\end{center}
\caption{Comparison of calculations and measurement for the deuteron
breakup coincidence cross section on $^{197}$Au at $E_d=12$~MeV
(Fig.~4 of \protect\cite{BaurRT84}). The postacceleration effect can
clearly be seen, as the maximum of the proton energy ($\sim 7.5$~MeV)
is larger than the one of the neutron ($\sim 2.5$~MeV).  The
experimental data are taken from \protect\cite{Jarczyk72}.
}
\label{fig:post}
\end{figure}

The present ``post-form'' description, Eqs.~(\ref{eq:postdwba}),
(\ref{eq:tintegral}) includes the effects of ``postacceleration''.
``Postacceleration'' arises in a purely classical picture of the
breakup process. This is nicely discussed in \cite{Galonsky94} (We
show their figure~5 here as our Fig.~\ref{fig:1}). The nucleus
$a=(c+n)$ moves up the Coulomb potential, loosing the appropriate
amount of kinetic energy. At the ``breakup point'' (marked as
``breakup occurs here'', see Fig~\ref{fig:1}), this kinetic energy
(minus the binding energy) is supposed to be shared among the
fragments according to their mass ratio (assuming that the velocities
of $c$ and $n$ are equal). Running down the Coulomb barrier, the
charged particle $c$ alone (and not the neutron) gains back the
Coulomb energy, resulting in its ``postacceleration''. Of course this
picture is based on the purely classical interpretation of this
process, and will be modified in a quantal treatment, where such a
``breakup point'' does not exist. The correct semiclassical limit of
the theory in this case can be found, e.g., in \cite{BaurPT74}. A
purely classical formula for this postacceleration, where the
``breakup point'' corresponds to the distance of closest approach ---
i.e., $b=r$ in Fig.~\ref{fig:1} --- is given in
\cite{BaurBK95}. Postacceleration is clearly observed in low energy
deuteron breakup, in the (fully quantal) theoretical calculations as
well as in the corresponding experiments, see Fig.~\ref{fig:post} and
also, e.g., \cite{BaurT76,BaurRT84}.

The formula Eq.~(\ref{eq:postdwba}) is also useful for the description
of the Coulomb dissociation of halo nuclei at high beam energies, see
\cite{ShyamBB92}. Within this theory postacceleration effects become
negligibly small in the high energy region. This is seen in the
numerical calculations \cite{ShyamBB92} and in the analytical
investigations to be described below. It can, e.g., be applied to
$^{11}$Be and $^{19}$C Coulomb dissociation experiments
\cite{Nakamura99,Nakamura94} (We disregard here the importance of
finite range effects).

On the other hand the 1$^{st}$ order semiclassical Coulomb excitation
theory was widely applied in the past years to the Coulomb
dissociation of high energy neutron halo nuclei, see, e.g.,
\cite{BaurHT01}.  The theory corresponds to the ``prior form'',
mentioned above.  The question of higher order electromagnetic effects
was studied recently in \cite{TypelB01} within this framework. These
effects were found to be small, for zero range as well as finite range
wave functions of the $a=(c+n)$-system. It seems interesting to note
that postacceleration effects arises through higher order
electromagnetic effects in straight line semiclassical theories, see
\cite{TypelB94}. Through the interference of 1$^{st}$ and 2$^{nd}$
order amplitudes even a ``post-deceleration'' can arise, as was seen
in that paper.

\begin{figure}[bt]
\begin{center}
\resizebox{0.3\textwidth}{!}{%
\includegraphics{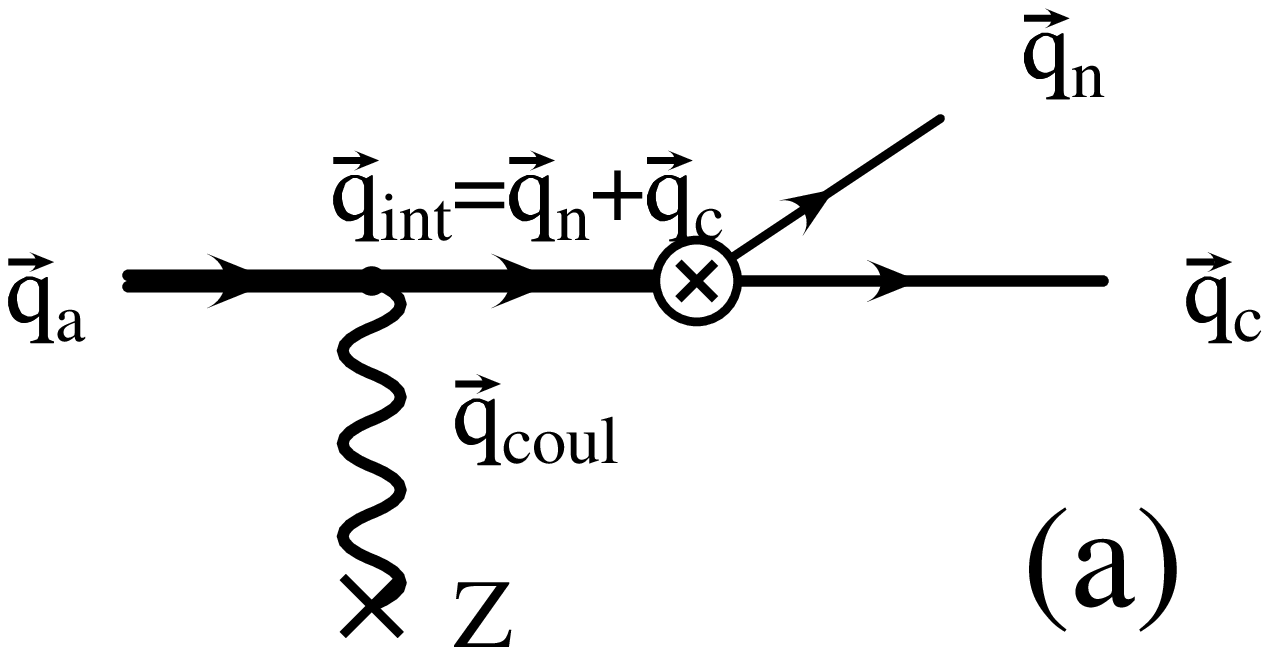}}
~
\resizebox{0.3\textwidth}{!}{%
\includegraphics{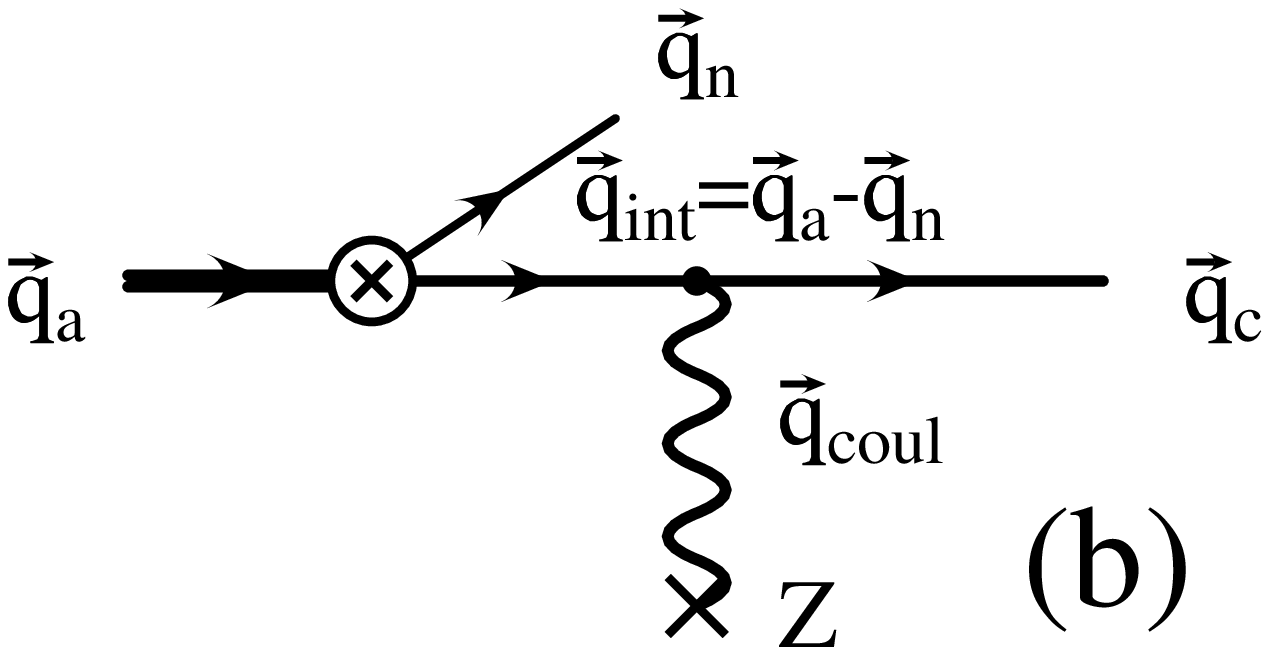}}
\end{center}
\caption{The two bremsstrahlung type of graphs, which describe the Coulomb
breakup in the Born approximation. Three-momentum conservation at each vertex
determines the intermediate momenta $\vec q_{\mathrm{int}}$.}
\label{fig:2}
\end{figure}

In this work we want to establish the relation between the apparently
very different post-form CWBA and semiclassical theory. It was
recently noticed \cite{BaurHT01} that in the limit of Coulomb
parameters $\eta_a=Z_c Z e^2 / \hbar v_a \ll 1$ (i.e. in the Born
approximation), where $v_a$ denotes the velocity of particle $a$
($v_a=\frac{\hbar q_a}{m_a}$), both theories give the same result.
Expanding the Coulomb wave functions up to first order in the Coulomb
fields one finds
\pagebreak
\begin{eqnarray}
T_{\mathrm{Born}} &=&  f_{\mathrm{coul}} D_0 \times \label{eq:born}\\
&&\left[ \frac{1}{q_a^2 - \left[\vec q_c + \vec q_n \right]^2}
+ \frac{m_n}{m_a \left[q_c^2 - \left(\vec q_n - \vec q_a\right)^2\right]}
\right]. \nonumber
\end{eqnarray}
Here $f_{\mathrm{coul}}=2\eta_a q_a/\left(\vec
q_{\mathrm{coul}}\right)^2$ is the usual Coulomb amplitude with the
``Coulomb push'' $\vec q_{\mathrm{coul}}=\vec q_a-(\vec q_c+\vec q_n)$, for
further details see \cite{BaurT72}.  The two terms in the parenthesis
correspond to the two graphs shown in Fig.~\ref{fig:2}. For small
values of $q_{\mathrm{coul}}$ the two terms almost cancel and the expansion in
$q_{\mathrm{coul}}$ was found to be in agreement with the semiclassical result,
see \cite{BaurHT01} and below.

We now show that this agreement is also true in the case of arbitrary
values of $\eta_a$ and $\eta_c$. The beam energy must be high
(compared to the binding energy $E_{bind}$) and the two fragment need
to be scattered into forward angles.

This is reminiscent of the result in the theory of
bremsstrahlung. Replacing the neutron by a photon the diagrams of
Fig.~\ref{fig:2} are identical to the bremsstrahlung in lowest
order. In this case it was already noticed that the Born result
remains valid {\em for arbitrary values of $\eta$} for high energies
and small scattering angles \cite{LandauQED}. Here we want to show
that the same applies in this case.

The $T$-matrix can be evaluated analytically in this model due to well
 known Nordsieck formula \cite{Nordsieck54}, see Eqs.~(11)--(13) of
 \cite{BaurT72}.  Using this formula one obtains the $T$-matrix
 Eq.~(\ref{eq:postdwba}) in terms of a hypergeometric function $F$ as
 well as its derivative $F'$. The argument of the hypergeometric
 function $F$ (and $F'$) is given by \cite{BaurT72,ShyamBB92}:
\begin{eqnarray}
&&\zeta(\lambda) = \\
&&\frac{2 q_{\mathrm{coul}}^2 (q_a q_c + \vec q_a \vec q_c)
-4(\vec q_{\mathrm{coul}} \vec q_a + \lambda q_a)(\vec q_{\mathrm{coul}} 
\vec q_c - \lambda q_c)}
{(q_{\mathrm{coul}}^2-2\vec q_{\mathrm{coul}} \vec q_a- 2 \lambda q_a)
(q_{\mathrm{coul}}^2+2\vec q_{\mathrm{coul}} \vec q_c - 2
\lambda q_c)}. \nonumber
\end{eqnarray}
We observe that (for $\lambda=0$) this parameter $\zeta(0)$ is found
to be negative and $-\zeta(0) \gg 1$ for beam energy large compared to
the binding energy and for perpendicular momentum transfers $q_\perp
\gg 2 \eta_a q_\|$ (nonadiabatic case), where $q_\| = \omega/v$ with
$\hbar \omega=E_{\mathrm{bind}} + E_{\mathrm{rel}}$ and where the
relative energy between $c$ and $n$ is $E_{\mathrm{rel}}=\frac{\hbar^2
q^2}{2 \mu}$ with the relative momentum given by $\vec q=\frac{m_c
\vec q_n - n_n \vec q_c}{m_a}$.  It was already noticed in the
numerical evaluation of the process that, due to $-\zeta(0)\gg1$ that
the hypergeometric series does not converge and an analytic
continuation had to be used. Here we use this fact to our advantage
and make a linear transformation to get the argument of the
hypergeometric function close to $0$. The transformation we are using
leads to the argument of the hypergeometric function $z=
\frac{1}{1-\zeta(0)}$ (Eq.~15.3.7 of \cite{AbramowitzS64}). In this
respect our approach differs from the one used in the bremsstrahlung
case, where a transformation giving an argument close to one is
used. Using only the lowest order term in the hypergeometric series
one obtains after some algebra (up to an overall phase)
\begin{eqnarray}
 T &\approx&  4\pi \ D_0 \ f_{\mathrm{coul}} \ e^{-\frac{\pi}{2}\xi} 
\biggl[
e^{-i\phi} \frac{1}{q_a^2-(\vec q_n + \vec q_c)^2} \label{eq:tdwba}\\
&& +
e^{+i\phi} \frac{m_c}{m_a}\ \frac{1}{q_c^2-(\vec q_n - \vec q_a)^2}
\biggr]. \nonumber
\end{eqnarray}
Hereby, the relative phase is $\phi = \sigma_0(\eta_c) -
\sigma_0(\eta_a) - \sigma_0(\xi) - \xi/2 \log |\zeta(0)|$. The
$\sigma_0(\eta)=\mathrm{arg}\Gamma(1+ i \eta)$ are the usual Coulomb
phase shifts, and $\xi=\eta_c-\eta_a$. The correspondence to the Born
result is clearly seen. One only has an additional prefactor
$e^{-\frac{\pi}{2}\xi}$ and a relative phase $e^{\pm i\phi}$ between
the two terms. The phase $\phi$ obviously is $O(\xi)$.  Since $v_c\sim
v_a$ the quantity $\xi$ is usually very small and so is $\phi$ for the
cases of \cite{Nakamura99,Nakamura94}. The prefactor is also well
known in the semiclassical theory, where it accounts for the
replacement of the ``Coulomb bended'' trajectories with the straight
line trajectories.  Both corrections vanish in the limit
$\xi\rightarrow 0 $ and the result coincides with the usual Born
approximation ({\em even if $\eta_a$ and $\eta_c$ are not small}).

We have seen that the $T$-matrix in the case of large Coulomb
parameters $\eta_a$ and $\eta_c$ corresponds to the Born result (small
Coulomb parameter) in the sudden (or nonadiabatic) case $q_\perp \gg 2
\eta_a q_\|$. We note that the derivation of Eq.~(\ref{eq:tdwba}) only
depends on the condition $-\zeta(0)\gg1$ (and not on the values of the
$\eta$'s).  For $\eta_a,\eta_c\gg1$ one can define a classical path
for both $a$ and $c$ in the initial and final state and
Eq.~(\ref{eq:tdwba}) can be related to the semiclassical approach (see
the discussion below following Eq.~(\ref{eq:lo})). We expect to find a
connection between the semiclassical theory and the adiabatic case
($q_\perp < 2 \eta_a q_\|$) and the fully quantal expression for the
$T$-matrix. For the adiabatic regime the well known exponential
decrease with the adiabacity parameter is observed in the numerical
calculations. In this case, the inequality $-\zeta(0)\gg1$ is not
generally satisfied. We are presently working to see how the
semiclassical limit can be obtained with analytical methods in this
case also.  Such a method would also be valid in both the adiabatic
and nonadiabatic case as long as $\eta_a,\eta_c\gg1$.

A similar situation is encountered in the theory of bremsstrahlung and
Coulomb excitation, see Section II~E of \cite{AlderBH56}.  There a
fully quantal expression for the differential cross-section for dipole
Coulomb excitation is given in II~E.62. It looks similar to the
corresponding expression for Coulomb breakup (see \cite{BaurT72}). The
semiclassical "variant" of this formula is found in II~E.57. It is
noted there that it can be obtained from the quantal expression by
letting at the same time $\eta_a$ and $\eta_c$ go to infinity and
perform a confluence in the hypergeometric functions.

\section{Scaling Properties}
\label{sec:2}
In many experimental situations the Coulomb push $\vec q_{\mathrm{coul}}$ 
is small. Having found that the full CWBA results agrees with the Born
result for small scattering angles, we can expand Eq.~(\ref{eq:born})
or Eq.~(\ref{eq:tdwba}) with $\phi=\xi=0$ for small values of
$q_{\mathrm{coul}}$. We obtain
\begin{eqnarray}
T &=& f_{\mathrm{coul}} \frac{2 D_0}{\pi^2} 
\frac{m_n^2 m_c}{m_a^3}
\frac{2 \vec q \cdot \vec q_{\mathrm{coul}}}
{\left(\kappa^2+q^2\right)^2}.
\label{eq:lo}
\end{eqnarray}
This result is in remarkable agreement with the usual 1$^{st}$ order
treatment of electromagnetic excitation in the semiclassical
approximation.

In the semiclassical approach the scattering amplitude is given by the
elastic scattering (Rutherford) amplitude times an excitation
amplitude $a(b)$, where the impact parameter is related to the
$q_\perp$ and $\eta$, see above. The absolute square of $a(b)$ gives
the breakup probability $P(b)$, in lowest order (LO).  It is given by
\cite{BaurHT01,TypelB01}
\begin{eqnarray}
\frac{dP_{\mathrm{LO}}}{dq} = \frac{16 y^2}{3\pi\kappa} 
\frac{x^4}{(1+x^2)^4},
\end{eqnarray}
where the variable $x$ is related to the relative momentum between $n$
and $c$ by $x=\frac{q}{\kappa}$ and $y$ is a strength parameter given
by
\begin{equation}
y = \frac{2 Z Z_c m_n e^2}{\hbar v_a m_a b \kappa}.
\end{equation} 

This formula shows very interesting scaling properties: Very many
experiments, for neutron halo nuclei with different binding energy,
beam energy, scattering angles (or $\vec q_n$ and $\vec q_c$) {\em all
lie on the same universal curve}!  (Corrections for finite values of
$\xi_{\mathrm{eff}}=\omega b/v =\xi(\theta)=2\eta_a q_\| /q_\perp $
should also be applied, according to \cite{TypelB01}.)  It will be
interesting to see in future calculations under what conditions (beam
energy, \dots) one finds deviations from this simple scaling
behavior. E.g., postacceleration effects will lead to such scaling
violations.

\section{Conclusion and  Outlook}
\label{sec:3}

The present model can be seen as a ``theoretical laboratory'', which
allows to study analytically, as well as, numerically the relation
between quantal and semiclassical theories, and the importance of
postacceleration effects.  We mention that from an experimental point
of view, the postacceleration effects are not fully clarified, see,
e.g., \cite{Nakamura94,Ieki93} ("postacceleration") and on the other
hand \cite{Bush98} ("no postacceleration").  Finally, let us mention
recent work on the electromagnetic dissociation of unstable
neutron-rich oxygen isotopes \cite{Leistenschneider01,Aumann01}.
These authors deduce photoneutron cross-sections from their
dissociation measurements.  If the neutrons are emitted in a slow
evaporation process in a later stage of the reaction, the question of
postacceleration is not there. On the other hand, for the light nuclei
there is some direct neutron emission component and the present kind
of theoretical analysis further proves the validity of the
semiclassical approach used in \cite{Leistenschneider01}.

Postacceleration effects are also of importance for the use of Coulomb
dissociation for the study of radiative capture reactions of
astrophysical interest. We expect that our present investigations will
shed light on questions of postacceleration and higher order effects
in these cases also.

\end{document}